# Circumnuclear starbursts in barred galaxies

Johan H. Knapen

Département de Physique, Université de Montréal, C.P. 6128, Succursale Centre-Ville, Montréal (Québec), H3C 3J7 Canada; and Observatoire du Mont Mégantic
*Present address:* Division of Physical Sciences, University of Hertfordshire, College Lane, Hatfield, Herts AL10 9AB, UK. E-mail knapen@star.herts.ac.uk

**Abstract.** I discuss[1] new observations of circumnuclear regions in barred galaxies, both of their kinematics and their near-infrared (NIR) morphology. New kinematic observations of M100 confirm that (1) the inner bar-like feature detected before in this galaxy is in fact a continuation of the large-scale stellar bar; and (2) the spiral arm-like sites of star formation in the circumnuclear zone are manifestations of density wave spiral structure in the core of this galaxy. From a NIR study of similar regions in other barred galaxies, I find that in many of these regions, sites of star formation are visible in the NIR images, implying that young stars are responsible for a non-negligible part of the NIR emission.

## 1 Introduction

Barred galaxies often experience starburst activity in their central regions, and in some cases the star formation (SF) is observed in more or less complete nuclear rings (see review by Kennicutt 1994). Imaging in the Hα emission line usually brings out these structures very well (e.g. Pogge 1989), but may be subject to dust extinction. This alone is sufficient reason to include near-infrared (NIR) imaging in the study of circumnuclear regions (CNRs) in barred galaxies.

In a recent study, we combined optical and NIR $K$-band imaging of the core of M100 (=NGC 4321), a mildly barred galaxy, with dynamical modelling (Knapen et al. 1995a,b; Shlosman 1996). The SF in the CNR of this galaxy delineates two inner spiral arms, flanked by dust lanes, which connect outward through the bar and to the main spiral arms in the disc. The circumnuclear SF occurs between a pair of inner Lindblad resonances (ILRs), as indicated by the NIR morphology and by the modelling. The morphology in Hα and blue light is strikingly different from that at 2.2 $\mu$m ($K$-band; Knapen et al. 1995a). In $K$, the spiral armlets are hardly discernable: the CNR is mostly smooth where the SF occurs. Closer to the centre though, an inner bar-like feature is seen in the NIR, along with two small leading arms (Knapen et al. 1995b; Shlosman 1996).

In the present paper, I first present new velocity observations which confirm kinematically the inner bar-like structure and the spiral nature of the SF regions in the core of M100. So far, these had only been inferred by us from optical and NIR imaging, and from dynamical modelling. I then discuss that the large difference between NIR and Hα core morphology as observed in M100 seems to be rather unusual compared to similar systems. From new high-resolution NIR imaging it is clear that in many, possibly most, cases of circumnuclear SF activity, the individual SF regions show up prominently in $K$ (unlike in M100).

## 2  Kinematics of the core of M100

### 2.1   CO

Recently, Rand (1995) observed parts of M100, including the central region, in CO with the BIMA interferometer. We re-analyzed some of his data in order to specifically study the gas kinematics in the core region. We first removed noise peaks outside the area where CO emission is expected by setting pixel values at those positions to undefined. This was done interactively by inspecting the individual channel maps one by one, continually comparing with the same and adjacent channels in a smoothed data cube. The resulting data set was used to calculate the total intensity and velocity moment maps. The resulting moment images show the same structure as published by Rand (1995) but are somewhat more sensitive in the central region. We show the velocity field thus produced, overlaid on a grey-scale representation of the total CO intensity, in Fig. 1.

Density wave streaming motions are recognizable in the deviations from the regular shape of the velocity contours (isovels), especially toward the NE and SW of the nucleus near radii of some 7″. These are due to the gas streaming near the spiral armlets (see also next section). In the central ∼ 5″, the isovels do not run parallel to the minor axis but show a deviation characteristic of gas streaming along a bar. Such deviations were described before by Bosma (1981) in several barred galaxies. Knapen et al. (1993) confirmed the existence of a weak bar component in M100 from the H I kinematics. Note, however, that the deviations seen here in CO occur on the much smaller scale of the inner bar-like feature as seen in the NIR (Knapen et al. 1995a), and confirm the gas streaming along this bar as seen in the numerical modelling (Knapen et al. 1995b; Shlosman 1996).

It is generally dangerous to use moment maps for interpretation of kinematical features in regions where profiles may deviate from a Gaussian shape, and/or be multiple-peaked. Since the inner region under consideration here is clearly such a region of enhanced risk, we produced a set of position-velocity diagrams along and parallel to the minor axis, shown in Fig. 2. It is immediately clear from this panel that multiple velocity components are present in the molecular gas, especially at the central position, and that the moment analysis may indeed not be valid there. But more interesting are 4 components which show up symmetrically in the position-velocity diagrams north and south of the minor axis. Two of these (labelled AN and AS in Fig. 2) are visible at RA offsets ∼ −8″ (North) and ∼ +8″ (S), with excess velocities with respect to $v_{sys}$ of ∼ +25 km s$^{-1}$ (N) and ∼ −25 km s$^{-1}$ (S). We identify these components as the density wave streaming motions near the spiral armlets seen before in the velocity field. Their offsets, both in RA and in velocity, strongly support this interpretation, as well as their symmetric occurrence in the series of L-V diagrams.

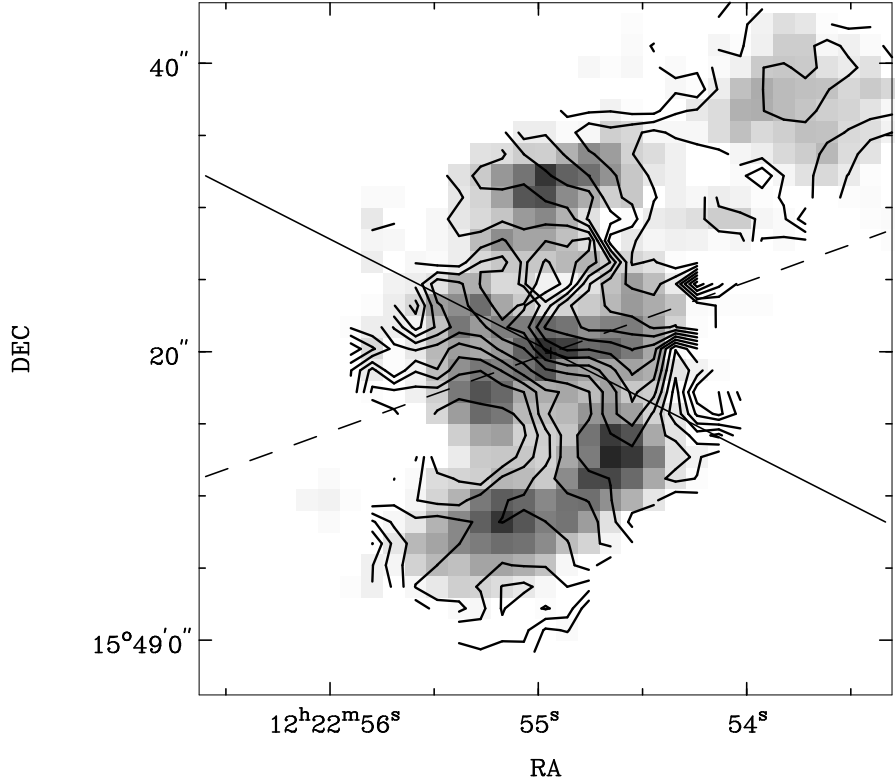

**Fig. 1.** CO velocity field of the central region of M100 in contours, overlaid on the total intensity CO map. Drawn line indicates kinematic minor axis, dashed line is position angle of the large-scale bar. Contours are separated by 10 km s$^{-1}$. Epoch is J2000.0. CO data from Rand (1995).

The second set of components (labelled BN and BS in Fig. 2) has RA offsets of $\sim +3''$ (N) and $\sim -3''$ (S), and excess velocities of $\sim +70$ km s$^{-1}$ (N) and $\sim -70$ km s$^{-1}$ (S). This is the gas streaming along the inner bar-like feature, as again indicated strongly by the symmetric offsets in both position and velocity, and by qualitative comparison with the velocity field. This kinematically observed gas streaming confirms the existence of the inner barlike feature (continuation of the outer bar) so prominently seen in our NIR imaging and dynamical modelling (Knapen et al. 1995a,b).

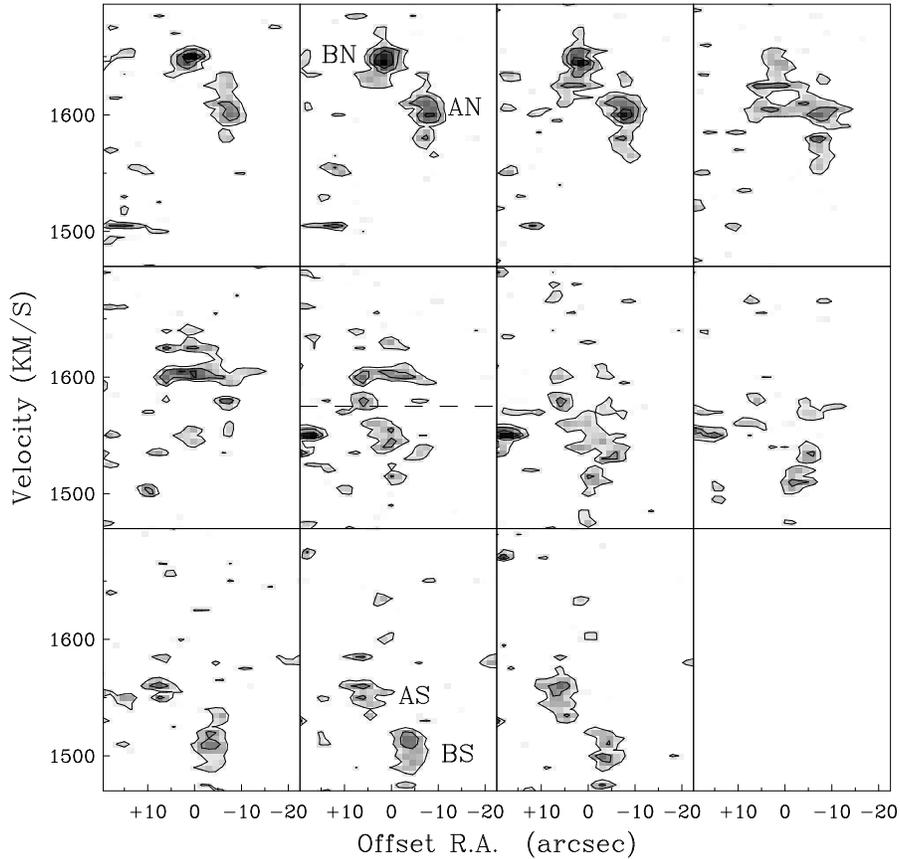

**Fig. 2.** Position-velocity diagrams in CO along (sixth panel) and parallel to the minor axis in the central region of M100. Panels to the left of and above the middle panel are cuts North of the minor axis, panels to the right of and below are South. Cuts are separated by  $1''5$ , or about half a beam. The systemic velocity of the galaxy is 1575  km s$^{-1}$, indicated by the horizontal line in the minor axis (middle) panel. Named features are discussed in the text.

It is interesting to compare Fig. 2 with Fig. 9 of Knapen et al. (1993). The latter are similar position-velocity diagrams along and parallel to the minor axis of M100, but show the H I kinematics on the scale of the large bar, which extends some 4 kpc in radius along a position angle of $\sim 110°$. The similarity between the gas components labelled BN and BS here, and the H I components with velocities symmetrically offset from $v_{sys}$ is striking. The H I behaviour was interpreted by Knapen et al. (1993) as streaming along the large-scale bar of the galaxy. The CO velocities similarly represent the streaming along the inner bar.

Both these kinematic observations indicate strongly that the gas is moving in the gravitational field of the same stellar bar.

## 2.2  Hα

Using the TAURUS instrument in Fabry-Perot (FP) mode on the 4.2m WHT on La Palma, we obtained a data set giving full 2-D velocity information in the Hα emission line. We observed two different parts of the disc of M100, and since the central region of the galaxy was present in both data sets, we combined these to obtain a core data cube with increased S/N. The data reduction procedure will be described in detail in a future publication. In the present paper, we present first results from a moment analysis in the central region of M100: the total intensity map and velocity field, shown in Fig. 3. The individual Hα profiles over all but the very innermost ($< 1''$) region are well suited for a moment analysis, double-peaked profiles being practically absent.

The total intensity map as derived from the FP observations is comparable in quality to the image published by Knapen et al. (1995a). The resolution is around $0\rlap{.}''7$ , indicating that the seeing was good and constant during the time of the FP observations ($> 6$ hrs). The velocity field of the central region shows rapidly increasing rotation velocities (the rotation curve as derived from the velocity field rises to more than $150 \,\mathrm{km\,s^{-1}}$ in $\sim 2''$, or some 150 pc), but also important deviations from circular velocities. The first effect to note specifically is the S-shaped deviation of the isovels near the minor axis, very similar to the deviation seen in CO, above. Also in Hα, the streaming of gas along the inner part of the bar shows up clearly in the velocities, confirming the earlier findings.

Another important deviation in the velocity field is seen most clearly toward the NE and SW of the nucleus, at radii of some $9''$. This deviation is due to density wave streaming motions, strongest where we inferred the position of the inner spiral arms, just outside the well-defined dust lanes (Knapen et al. 1995a,b). Although the signature of the streaming motions is most obviously visible near the minor axis of the galaxy, they can in fact be recognized consistently up to some 60° on either side of the minor axis. As estimated from the velocity contours, the (projected) excess or streaming velocities are of the order of $40 \,\mathrm{km\,s^{-1}}$. This kinematic detection confirms that indeed the inner spiral armlets are miniature density wave spiral arms, with a behaviour very similar to what one is used to see in discs of galaxies (e.g. Visser 1980).

## 3   NIR imaging of CNRs in barred galaxies

The morphology of the central 2 kpc region in M100 changes most strikingly from the blue to the NIR (Knapen et al. 1995a,b). Whereas in the blue, and also in Hα, the core region is dominated by a pair of miniature spiral arms which show strong SF activity, the $K$ contours in the region are remarkably smooth. They hardly show any structure apart from two symmetrically placed "hot spots".

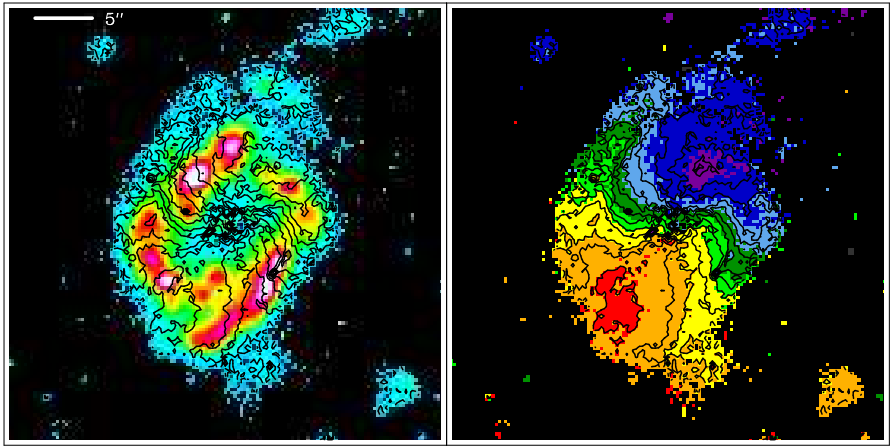

**Fig. 3.** Hα FP moment maps of the central region of M 100: the velocity field is shown in contours and overlaid on the total intensity Hα image (left), and on the velocity field itself (right). Contours are separated by 15 km s$^{-1}$. N is up, E to the left.

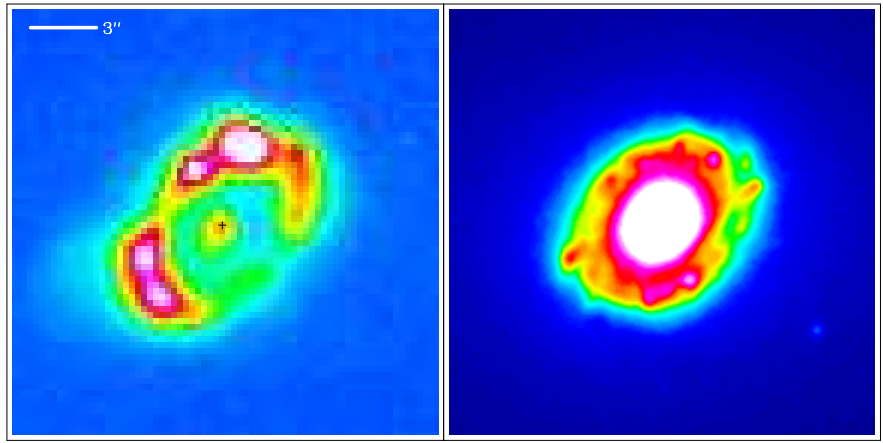

**Fig. 4.** Images of the core of NGC 6951 in Hα (left) and in the NIR $K$-band (right). The two images are at the same scale and orientation. N is up, E to the left. The bar runs almost horizontally.

From a recent NIR imaging survey of a number of systems with M100-like Hα morphology, it becomes clear that M100 is more exception than rule. As an example, we show in Fig. 4 an Hα (left, taken with TAURUS on the WHT) and a $K$ image (right, from MONICA on the CFHT), of the inner $\sim 20''$ region of the strongly barred Seyfert galaxy NGC 6951. Unlike in M100, the $K$ image of NGC 6951 shows a number of distinct sites of enhanced emission in the radial region where the intense SF occurs, at some $2'' - 4''$ from the nucleus. There is a general correspondence between the positions of the "hot spots" in Hα and in $K$, although some regions emit relatively more in Hα, others more in $K$. In some cases the Hα is seen slightly offset from a $K$-peak (e.g. the strongest $K$ feature toward the SE, near which a strong dust lane can be seen in NIR colour index images).

From our mini-survey, we find that most of the studied objects are in fact similar to NGC 6951 in that the $K$ images of the CNRs show a number of distinct emitting regions, often coinciding spatially with concentrations of Hα emission. Only a few of the objects studied show M100-like smooth $K$ emission in SF regions which are so prominent in Hα. Additional observational data is necessary to explain this difference between M100 and other galaxies in our sample. We note, however, that one expects a region of massive SF to become prominent in $K$ after some $10^7$ yrs, when the massive stars reach the supergiant phase in their evolution (e.g. Leitherer & Heckman 1995). This implies that the observations of NGC 6951 can be explained rather straightforwardly if one assumes that bursts of SF of relatively short duration (around $10^7$ yrs) occur randomly in the CNR. Then some regions emit more in Hα (starburst younger than $10^7$ yrs, no supergiants yet) and others more in $K$ (slightly older, no more current SF). This interpretation implies that something prevents the SF regions from shining in $K$ in M100-like systems, unless the entire CNR is younger than $10^7$ yrs, which is unlikely. A more reasonable explanation is that the newly formed stars are dispersed throughout the CNR, so that by the time they become supergiants they are no longer as clustered as at birth (see Knapen et al. 1995a,b). This idea awaits further observational confirmation.

## 4   Discussion

The results of the $K$-band imaging (Sect. 3) seem to bring up more questions than they answer, and bring out the need for further detailed observations. Specifically, what needs to be addressed is what kinds of stars, and in which proportion, contribute to the NIR emission from the central regions of barred spirals. Because one often observes localized $K$-band emission from SF regions, at least in those cases an old stellar population cannot be entirely responsible for the $K$-emission in the central regions, and emission from young, massive stars must play a non-negligible role. There are several possible ways to quantify the relative contributions of old and young stars (basically: to disentangle the giant and supergiant contributions) to the NIR emission at the scales of interest here.

One is detailed spectroscopy in the NIR (e.g. Goldader et al. 1995), another imaging in the NIR $J, H$ and $K$ bands and across the CO $2.3\mu m$ absorption feature (e.g. Forster 1994; Armus et al. 1995; Haller et al. 1996).

Another open question is why in some galaxies the $K$ light is smoothly distributed in the CNR, even though the SF as seen in H$\alpha$ occurs in discrete regions (e.g. M100: Knapen et al. 1995a), whereas in others the $K$ follows the H$\alpha$ emission, thus both trace the SF (e.g. NGC 6951: Fig. 4). The difference must be due to either different stellar populations in the two classes of galaxies, or to different dynamics. Two-dimensional kinematic observations can be used to study the latter possibility. It is interesting to note that in the case of NGC 6951 the velocity field as derived from H$\alpha$ FP observations (Knapen, in preparation) is regular in the radial region where the massive SF is concentrated, and does not show density wave streaming motions, certainly not of the amplitude seen in M100. Possibly the CNR of NGC 6951 is older and the mini-spiral is much more tightly wound, resembling a complete ring where the SF can occur (see e.g. Elmegreen 1994). In the latter case, stars will evolve more or less at the place where they formed, and H$\alpha$ (tracing very young massive stars) and $K$ (same stars after $\sim 10^7$ yrs) emission spatially coincide. In the former case (M100), stars may be displaced from their birthplace during the $10^7$ yrs of their evolution to supergiants, and dispersed enough to smooth their emission in the $K$-image, due to a sheared motion in the disc. Further observations will clarify this point.

ACKNOWLEDGEMENTS I wish to thank the organizers of this symposium for making it possible for me to attend it. I am indebted to my collaborators on the various aspects of the work described in this paper: J.E. Beckman, R. Doyon, C.H. Heller, R.S. de Jong, D. Nadeau, R.F. Peletier, R.J. Rand, M. Rozas, and I. Shlosman. The WHT is operated on the island of La Palma by the RGO in the Spanish Observatorio del Roque de los Muchachos of the IAC. The CFHT is operated by the NRC of Canada, the CNRS of France and the Univ. of Hawaii.